\documentclass[aps,prd,onecolumn,groupedaddress,showpacs,nofootinbib,amssymb]{revtex4}
\usepackage[dvips]{graphicx}
\usepackage{amssymb}
\usepackage{amsmath}
\usepackage{graphicx,color}
\usepackage{amsfonts}
\usepackage{bm}
\usepackage{cancel}
\usepackage{comment}
\usepackage{hyperref}
\usepackage{ulem}

\newcommand{\e}{\mathrm{e}} 

\allowdisplaybreaks[4]

\begin{document}

\tolerance=5000

\title{General geometry realized by four-scalar model and application to $f(Q)$ gravity}

\author{G.~G.~L. Nashed$^{1}$}\email{nashed@bue.edu.eg}
\author{Shin'ichi~Nojiri$^{2,3}$}\email{nojiri@gravity.phys.nagoya-u.ac.jp}

\affiliation{ $^{1)}$ Centre for Theoretical Physics, The British University in Egypt, P.O. Box
43, El Sherouk City, Cairo 11837, Egypt \\
$^{2)}$ Department of Physics, Nagoya University, Nagoya 464-8602, Japan \\
$^{3)}$ Kobayashi-Maskawa Institute for the Origin of Particles and the Universe, Nagoya University, Nagoya 464-8602}

\begin{abstract}

In this paper, we propose a model including four scalar fields coupled with general gravity theories, which is a generalization of the two-scalar model
proposed in Phys. Rev. D \textbf{103} (2021) no.4, 044055, where it has been shown that any given spherically symmetric static/time-dependent spacetime
can be realized by using the two-scalar model.
We show that by using the four-scalar model, we can construct a model that realizes any given spacetime as a solution even if the spacetime does not have
a spherical symmetry or any other symmetry.
We also show that by imposing constraints on the scalar fields by using the Lagrange multiplier fields, the scalar fields become non-dynamical and they do not propagate.
This tells that there does not appear any sound which is usually generated by the density fluctuation of the fluid.
In this sense, the model with the constraints is a most general extension of the mimetic theory in JHEP \textbf{11} (2013), 135,
where there appears an effective dark matter.
The dark matter is non-dynamical and it does not collapse even under gravitational force.
Our model can be regarded as a general extension of any kind of fluid besides dark matter.
We may consider the case that the potential of the scalar fields vanishes and the model becomes a non-linear $\sigma$ model.
Then our formulation gives a mapping from the geometry of the spacetime to the geometry of the target space of the non-linear $\sigma$ model
via gravity theory although the physical meaning has not been clear.
We also consider the application of the model to $f(Q)$ gravity theory, which is based on a non-metricity tensor and $Q$ is a scalar quantity constructed from the non-metricity tensor.
When we consider the $f(Q)$ gravity in the coincident gauge where the total affine connections vanish,
when $f(Q)$ is not a linear function of $Q$, spherically symmetric spacetime cannot be realized except in the case that $Q$ is a constant.
The situation does not change if we use the two-scalar model, as we show.
If we use the four-scalar model in this paper, however, spherically symmetric spacetime can be realized
in the framework of $f(Q)$ gravity with the coincident gauge.

\end{abstract}

\maketitle

\section{Introduction}
\label{introduction}

When we consider the spacetime structure in any gravity theory, we solve the equations given by the model.
Inversely it is often, however, interesting to find a model that realizes the geometry desired from the theoretical and/or observational viewpoints.
After that, we can investigate whether the obtained model can be realistic or not.
We may call the systematic formulation ``reconstruction''.
In the case of the Friedmann-Lema\^{i}tre-Robertson-Walker (FLRW) spacetime, the formulation of the reconstruction has been well-studied
for the scalar-tensor theories~\cite{Nojiri:2005pu, Capozziello:2005tf}, the Einstein--scalar--Gauss-Bonnet gravity~\cite{Nojiri:2006je},
and $F\left(\mathcal{G}\right)$ gravity~\cite{Nojiri:2005jg, Nojiri:2005am, Cognola:2006eg, Nojiri:2019dwl},
where $\mathcal{G}$ is the Gauss-Bonnet topological invariant,
and $F(R)$ gravity~\cite{Nojiri:2006gh, Nojiri:2009kx}.
For the review of the modified gravity theories, see~\cite{Sotiriou:2008rp, Capozziello:2011et, Nojiri:2010wj, Nojiri:2017ncd}.

The reconstruction of the spherically symmetric spacetime has been also investigated by using two-scalar fields~\cite{Nojiri:2020blr}
and in the scalar--Einstein--Gauss-Bonnet gravity~\cite{Nashed:2021cfs}.
These models, however, include ghosts in general.
In the classical theory, the kinetic energy of the ghosts is unbounded below and there occur instabilities of the system.
In the quantum theory, the ghosts generate the negative norm states typically as in the Fadeev-Popov ghosts in the gauge theories~\cite{Kugo:1979gm}.
The negative norm states mean that the states have negative probabilities, which violates the Copenhagen interpretation of the quantum theory.
Therefore the existence of the ghost tells that the model is physically inconsistent.
It has been shown that the ghost can be eliminated by using constraints given by the Lagrange multiplier fields.
The constraints are similar to the mimetic constraint~\cite{Chamseddine:2013kea}, where non-dynamical dark matter appears.
By using the formulation, the models that realize stable wormholes~\cite{Nojiri:2023dvf}, gravastars with large surface redshift~\cite{Nojiri:2023zlp},
dynamical wormholes~\cite{Elizalde:2023rds} in the framework of the scalar--Einstein--Gauss-Bonnet gravity,
exotic objects where the stellar radius is less than the Schwarzschild one~\cite{Nojiri:2023ztz},
and wormholes inside stars and black holes~\cite{Nojiri:2024dde}.

In this paper, we consider more generalizations of the two-scalar model to the model with four scalar fields.
One motivation is in the $f(Q)$ gravity theory.
The $f(Q)$ gravity theory is not based on the curvature of spacetime nor torsion but on non-metricity tensor
and $f(Q)$ is a function of $Q$ which is a scalar quantity constructed from the non-metricity tensor.
In the case$f(Q) \propto Q + \mathrm{const}$, the theory reduces to Einstein's general relativity and therefore the model is called
the Symmetric Teleparallel Equivalent of General Relativity (STEGR).
In spite that there are many application of $f(Q)$ gravity~\cite{BeltranJimenez:2017tkd, BeltranJimenez:2019tme, Esposito:2021ect, DAmbrosio:2021pnd, Gadbail:2022jco,
Albuquerque:2022eac, Capozziello:2022wgl, Khyllep:2022spx, Anagnostopoulos:2022gej, Mustafa:2021bfs, Iosifidis:2018diy, Dimakis:2022rkd, Capozziello:2024vix},
the structure of $f(Q)$ gravity where $f(Q)$ is non-linear has not been well understood.
Especially the number of the physical degrees of freedom in this theory is under arguments~\cite{Hu:2022anq, DAmbrosio:2023asf, Hu:2023gui}.
The degrees of freedom have been often investigated in the coincident gauge where the total affine connections vanish
but the basis of the coincident gauge has not been so clear.
When $f(Q)$ is a non-linear function of $Q$, the spherically symmetric spacetime cannot be realized except in the case that $Q$ is a 
constant~\cite{Banerjee:2021mqk, Lin:2021uqa, Wang:2021zaz, Hohmann:2019fvf, DAmbrosio:2021zpm, Bahamonde:2022zgj, Zhao:2021zab} in the coincident gauge.\footnote{
If we do not use the coincident gauge, we may construct spherically symmetric spacetime like wormholes~\cite{DeFalco:2023twb}.
which we may detect observationally the wormholes by developing some astrophysical techniques~\cite{DeFalco:2021klh, DeFalco:2021btn, Zhao:2021zab}. }
We should note that in general, we may be able to choose the coordinate system consistent with the coincident gauge~\cite{Hohmann:2019fvf, DAmbrosio:2021zpm}. 
As we show, even if we use the two-scalar model, we cannot realize a spherically symmetric spacetime
in the framework of $f(Q)$ gravity with the coincident gauge.
If we use the four-scalar model proposed in this paper, spherically symmetric spacetime can be realized
in the framework of $f(Q)$ gravity with the coincident gauge.

Another strong motivation why we consider the four-scalar model is that we can reconstruct a model that realizes any given geometry, which is
time-dependent and not spherically symmetric in general.
The four-scalar model has potential in general but any spacetime can be reconstructed without potential.
When the potential vanishes, the four-scalar model can be regarded with a non-linear $\sigma$ model, whose target space is a four-dimensional manifold with geometry.
Therefore the formulation in this paper gives a mapping from the geometry of the real spacetime into the geometry of the target space in the $\sigma$ model.
There might be any deep meaning in the mapping, especially when we consider the quantum theory of gravity.

In the next section, we review the two-scalar model and find a condition for the gravity theory that any spherically symmetric spacetime can be realized.
In Section~\ref{fourscalar}, we propose a four-scalar model and show that any given geometry of the spacetime can be realized in this model.
We also show that the ghost can be eliminated by the constraints.
In Section~\ref{fQ}, we investigate the $f(Q)$ gravity and we show that general spherically symmetric spacetime cannot be realized by the two-scalar model but the four-scalar model
can do.
The last section is devoted to summary and discussion.

As conventions, the Greek indices $\mu,\nu=0,1,2,3$ and the Lattin indices $i,j=1,2,3$.
We also choose the signature of the metric as $(-,+,+,+)$. 

\section{Gravity coupled to two scalars}
\label{twoscalar}

In the framework of Einstein's gravity, we can realize arbitrarily given spherically symmetric spacetimes by using the model
where two scalar fields are coupled with gravity~\cite{Nojiri:2020blr}.
The model of \cite{Nojiri:2020blr}, however, includes ghosts, which make the model physically unacceptable.
This is because as a classical theory, the kinetic energy of the ghost is unbounded below, and as a quantum theory,
ghosts generate negative norm states, which conflicts with the Copenhagen interpretation \cite{Kugo:1979gm}.
After the paper~\cite{Nojiri:2020blr}, several works~\cite{Nojiri:2023dvf, Nojiri:2023zlp, Elizalde:2023rds, Nojiri:2023ztz} have shown
that the ghosts can be eliminated by constraints given by the Lagrange multiplier fields and the ghosts do not propagate..
These constraints can be regarded as a generalization of the mimetic constraint in \cite{Chamseddine:2013kea},
where non-dynamical dark matter effectively appears.
In this section, we consider general gravity theory minimally coupled with the two scalar fields.

We consider the following action as a generalization of the model in \cite{Nojiri:2020blr} including two scalar fields $\phi$ and $\chi$,
\begin{align}
\label{I8B}
S =&\, S_\mathrm{gravity} + S_{\phi\chi} + S_\mathrm{matter} \, , \nonumber \\
S_{\phi\chi} =&\, \int d^4 x \sqrt{-g} \left\{
 - \frac{1}{2} A (\phi,\chi) \partial_\mu \phi \partial^\mu \phi
 - B (\phi,\chi) \partial_\mu \phi \partial^\mu \chi
 - \frac{1}{2} C (\phi,\chi) \partial_\mu \chi \partial^\mu \chi
 - V (\phi,\chi) \right\}\, .
\end{align}
Here $S_\mathrm{gravity}$ is the action of the arbitrary gravity theory, $S_\mathrm{matter}$ is the action of matter,
$A(\phi,\chi)$, $B(\phi,\chi)$, and $C(\phi,\chi)$ are functions of $\phi$ and $\chi$,
and $V(\phi,\chi)$ is the potential for $\phi$ and $\chi$. 

By the variation of the action \eqref{I8B} with respect to the metric $g_{\mu\nu}$, we obtain the following
equation corresponding to the Einstein equation,
\begin{align}
\label{gb4bD4}
\mathcal{G}_{\mu\nu} = &\, \frac{1}{2} g_{\mu\nu} \left\{
 - \frac{1}{2} A (\phi,\chi) \partial_\rho \phi \partial^\rho \phi
 - B (\phi,\chi) \partial_\rho \phi \partial^\rho \chi
 - \frac{1}{2} C (\phi,\chi) \partial_\rho \chi \partial^\rho \chi - V (\phi,\chi)\right\} \nonumber \\
&\, + \frac{1}{2} \left\{ A (\phi,\chi) \partial_\mu \phi \partial_\nu \phi
+ B (\phi,\chi) \left( \partial_\mu \phi \partial_\nu \chi
+ \partial_\nu \phi \partial_\mu \chi \right)
+ C (\phi,\chi) \partial_\mu \chi \partial_\nu \chi \right\}
+ \frac{1}{2} \left( T_\mathrm{matter} \right)_{\mu\nu} \, .
\end{align}
Here $\left( T_\mathrm{matter} \right)_{\mu\nu}$ is the energy-momentum tensor of matter
defined by $\left( T_\mathrm{matter} \right)^{\mu\nu}\equiv \frac{2}{\sqrt{-g}} \frac{\delta S_\mathrm{matter}}{\delta g_{\mu\nu}}$ 
and $\mathcal{G}_{\mu\nu}$ is defined by the variation of the action $S_\mathrm{gravity}$ of the gravity sector as follows,
\begin{align}
\label{mathcalG}
\mathcal{G}^{\mu\nu} \equiv - \frac{1}{\sqrt{-g}} \frac{\delta S_\mathrm{gravity}}{\delta g_{\mu\nu}}\, .
\end{align}
Furthermore by the variations of the action \eqref{I8B} with respect to the two fields $\phi$ and $\chi$, we obtain
\begin{align}
\label{I10}
0 =&\, \frac{1}{2} A_\phi \partial_\mu \phi \partial^\mu \phi
+ A \nabla^\mu \partial_\mu \phi + A_\chi \partial_\mu \phi \partial^\mu \chi
+ \left( B_\chi - \frac{1}{2} C_\phi \right)\partial_\mu \chi \partial^\mu \chi
+ B \nabla^\mu \partial_\mu \chi - V_\phi \, , \nonumber \\
0 =&\, \left( - \frac{1}{2} A_\chi + B_\phi \right) \partial_\mu \phi \partial^\mu \phi
+ B \nabla^\mu \partial_\mu \phi
+ \frac{1}{2} C_\chi \partial_\mu \chi \partial^\mu \chi + C \nabla^\mu \partial_\mu \chi
+ C_\phi \partial_\mu \phi \partial^\mu \chi - V_\chi \, ,
\end{align}
Here $A_\phi=\partial A(\phi,\chi)/\partial \phi$, etc.
These field equations are nothing but the Bianchi identities if the equation $\nabla^\mu \mathcal{G}_{\mu\nu}=0$ is satisfied.

The metric of a general spherically symmetric and time-dependent spacetime is given by \cite{Nojiri:2020blr},
\begin{align}
\label{GBiv}
ds^2 = - \e^{2\nu (t,r)} dt^2 + \e^{2\lambda (t,r)} dr^2 + r^2 \left( d\vartheta^2 + \sin^2\vartheta d\varphi^2 \right)\, .
\end{align}
Without any loss of generality, it may be assumed as
\begin{align}
\label{TSBH1}
\phi=t\, , \quad \chi=r\, .
\end{align}
See \cite{Nojiri:2020blr, Nojiri:2023dvf, Nojiri:2023zlp, Elizalde:2023rds, Nojiri:2023ztz} for the detailed arguments.

As discussed in \cite{Nojiri:2020blr, Nojiri:2023dvf, Nojiri:2023zlp, Elizalde:2023rds, Nojiri:2023ztz}, 
$A$ and/or $C$ become often negative, and therefore $\phi$ and/or $\chi$ become ghosts.
The elimination of the ghosts can be done by imposing constraints by introducing the Lagrange multiplier fields $\lambda_\phi$ and $\lambda_\chi$.
We modify the action $S_{\phi\chi}$ in (\ref{I8B}) by $S \to S + S_\lambda$, where
\begin{align}
\label{lambda1}
S_\lambda = \int d^4 x \sqrt{-g} \left[ \lambda_\phi \left( \e^{2\nu(t=\phi, r=\chi)} \partial_\mu \phi \partial^\mu \phi + 1 \right)
+ \lambda_\chi \left( \e^{2\lambda(t=\phi, r=\chi)}\partial_\mu \chi \partial^\mu \chi - 1 \right) \right] \, .
\end{align}
The variations of $S_\lambda$ with respect to $\lambda_\phi$ and
$\lambda_\chi$ give the following constraints consistent with (\ref{TSBH1}),
\begin{align}
\label{lambda2}
0 = \e^{2\nu(t=\phi, r=\chi)} \partial_\mu \phi \partial^\mu \phi + 1 \, , \quad
0 = \e^{2\lambda(t=\phi, r=\chi)} \partial_\mu \chi \partial^\mu \chi - 1 \, .
\end{align}
Due to the constraints in Eq.~(\ref{lambda2}), the two scalar fields $\phi$ and $\chi$ become non-dynamical and the fluctuation of
$\phi$ and $\chi$ from the background given by (\ref{TSBH1}) do not propagate.
The fluctuations can be expressed as,
\begin{align}
\label{pert1}
\phi=t + \delta \phi \, , \quad \chi=r + \delta \chi\, .
\end{align}
By using Eq.~(\ref{lambda2}), we find
\begin{align}
\label{pert2}
\partial_t \left( \e^{2\nu(t=\phi, r=\chi)} \delta \phi \right) = \partial_r \left( \e^{2\lambda(t=\phi, r=\chi)} \delta \chi \right) = 0\, .
\end{align}
Eq.~(\ref{pert2}) tells that by imposing the initial condition $\delta\phi=0$ (because $\phi$ corresponds to the time coordinate)
and by imposing the boundary condition $\delta\chi\to 0$ when $r\to \infty$ (because $\chi$ corresponds to the radial coordinate),
we find that both of $\delta \phi$ and $\delta \chi$ vanish in the whole spacetime, $\delta\phi=0$ and $\delta\chi=0$
not only at the initial surface for $\delta\phi$ and boundary surface for $\delta\chi$. 
This tells that both $\phi$ and $\chi$ are non-dynamical or frozen degrees of freedom.

Refs.~\cite{Nojiri:2023dvf, Nojiri:2023zlp, Elizalde:2023rds, Nojiri:2023ztz} tell that $\lambda_\phi=\lambda_\chi=0$
consistently appear as a solution even in the model with the modified action $S + S_\lambda$.
This tells any solution of Eqs.~(\ref{gb4bD4}) and (\ref{I10}) which are based on the original action (\ref{I8B}) is
a solution even for the modified model with the action $S + S_\lambda$.

We now explain how we can construct a model that has a solution realizing given $\e^{2\nu(t,r)}$ and $\e^{2\lambda(t,r)}$ in (\ref{GBiv}).

The $(t,t)$, $(r,r)$, $(i,j)$, and $(t,r)$ components of (\ref{gb4bD4}) have the following forms,
\begin{align}
\label{TSBH2B}
2\mathcal{G}_{tt} =&\, - \e^{2\nu} \left( - \frac{A}{2} \e^{-2\nu} - \frac{C}{2} \e^{-2\lambda} - V \right) + \e^{2\nu} \rho \, , \nonumber \\
2\mathcal{G}_{rr} =&\, \e^{2\lambda} \left( \frac{A}{2} \e^{-2\nu} + \frac{C}{2} \e^{-2\lambda} - V \right) + \e^{2\lambda} p \, , \nonumber \\
2\mathcal{G}_{\vartheta \vartheta } = \frac{2\mathcal{G}_{\varphi\varphi}}{\sin^2 \vartheta}
=&\, r^2 \left( \frac{A}{2} \e^{-2\nu} - \frac{C}{2} \e^{-2\lambda} - V \right) + r^2 p \, , \nonumber \\
2\mathcal{G}_{tr} = 2\mathcal{G}_{rt} =&\, B \, , \nonumber \\
0=&\, \mathcal{G}_{t\vartheta } = \mathcal{G}_{\vartheta  t} = \mathcal{G}_{r\vartheta } = \mathcal{G}_{\vartheta  r}
= \mathcal{G}_{t\varphi } = \mathcal{G}_{\varphi  t} = \mathcal{G}_{r\varphi } = \mathcal{G}_{\varphi  r}\, .
\end{align}
Here we assume that the matter is a perfect fluid and $\rho$ and $p$ are the energy density and the pressure of matter, defined by,
\begin{align}
\label{FRk2}
\left(T_\mathrm{matter}\right)_{tt}=\rho=-g_{tt}\rho\ ,\quad \left( T_\mathrm{matter}\right)_{ij} =pg_{ij}\, .
\end{align}
In the case of Einstein's gravity, we find that $\mathcal{G}_{t\vartheta }=\mathcal{G}_{\vartheta  t}$,
$\mathcal{G}_{r\vartheta }=\mathcal{G}_{\vartheta  r}$,
$\mathcal{G}_{t\varphi }=\mathcal{G}_{\varphi  t}$, and $\mathcal{G}_{r\varphi }=\mathcal{G}_{\varphi  r}$ vanish trivially.
As we will see later, however, in the case of $f(Q)$ gravity $\mathcal{G}_{r\vartheta }=\mathcal{G}_{\vartheta  r}$ does not vanish and
the equation $\mathcal{G}_{r\vartheta }=\mathcal{G}_{\vartheta  r}=0$ gives non-trivial constraints on the model.

When $\mathcal{G}_{t\vartheta }=\mathcal{G}_{\vartheta  t}$, $\mathcal{G}_{r\vartheta }=\mathcal{G}_{\vartheta  r}$,
$\mathcal{G}_{t\varphi }=\mathcal{G}_{\varphi  t}$, and $\mathcal{G}_{r\varphi }=\mathcal{G}_{\varphi  r}$ vanish trivially,
the equations (\ref{TSBH2B}) can be algebraically solved with respect to $A$, $B$, $C$. and $V$ as follows,
\begin{align}
\label{ABCV}
& A= 2\left( \mathcal{G}_{tt} + \frac{\e^{2\nu}}{r^2} \mathcal{G}_{\vartheta \vartheta } \right) - \e^{2\nu} \left( \rho + p \right) \, , \quad
B= 2 \mathcal{G}_{tr} \, , \quad
C= 2\left( \mathcal{G}_{rr} - \frac{\e^{2\lambda}}{r^2} \mathcal{G}_{\vartheta \vartheta } \right) \, , \nonumber \\
& V= \left( \e^{-2\nu} \mathcal{G}_{tt} - \e^{-2\lambda} \mathcal{G}_{rr} \right) - \frac{1}{2} \left( \rho - p \right) \, .
\end{align}
Then we can obtain a model that realizes the spacetime defined by (\ref{GBiv}) by finding $(t,r)$-dependence of $\rho$ and $p$ and
by replacing $(t,r)$ in (\ref{ABCV}) with $(\phi,\chi)$.

\section{Four-scalar model}
\label{fourscalar}

Because in the gravity theory where any of $\mathcal{G}_{t\vartheta }=\mathcal{G}_{\vartheta  t}$,
$\mathcal{G}_{r\vartheta }=\mathcal{G}_{\vartheta  r}$,
$\mathcal{G}_{t\varphi }=\mathcal{G}_{\varphi  t}$, and $\mathcal{G}_{r\varphi }=\mathcal{G}_{\varphi  r}$
does not vanish, even if we use the two-scalar model in
Section~\ref{twoscalar}, we cannot construct a model which realizes a given spherically symmetric spacetime.
In order to solve the problem, we now consider further extension of the model by using four scalar fields.
We also show that we can realize any geometry by using the model.

We now consider the following action including four scalar fields $\phi^{\left(\rho\right)}$ $\left(\rho=0,1,2,3\right)$,
\begin{align}
\label{acg1}
S =&\, S_\mathrm{gravity} + S_\phi + S_\lambda \, , \nonumber \\
S_\phi \equiv&\, \int d^4x \sqrt{-g} \left\{ \frac{1}{2} \sum_{\rho,\sigma=0,1,2,3} A_{\left(\rho\sigma\right)} \left(\phi^{\left(\tau\right)}\right)
g^{\mu\nu} \partial_\mu \phi^{\left(\rho\right)} \partial_\nu \phi^{\left(\sigma\right)} - V\left( \phi^{\left(\rho\right)} \right) \right\}\, , \nonumber \\
S_\lambda \equiv&\, \int d^4x \sqrt{-g} \sum_{\rho=0,1,2,3} \lambda^{(\rho)} \left( \frac{1}{g^{\rho\rho}\left( x^\sigma = \phi^{\left( \sigma \right)} \right)} g^{\mu\nu} \left( x^\sigma \right)
\partial_\mu \phi^{\left(\rho\right)} \partial_\nu \phi^{\left(\rho\right)} - 1\right) \, .
\end{align}
Here the kinetic coefficients $A_{\left(\rho\sigma\right)} \left(\phi^{\left(\tau\right)}\right) $
and the potential $V\left( \phi^{\left(\rho\right)} \right)$ are functions of the scalar fields $\phi^{\left(\rho\right)}$.
We should note that the indices $\rho\sigma$ in $A_{(\rho\sigma)}$ etc. do not represent the indices of the tensor. 
In order to distinguish them, we use the notation with parentheses. 
We used Greek indices, however, because we will identify $\phi^{(\rho)}$ with $x^\rho$ later on. 
For simplicity, we exclude the contributions from matter since their inclusion, while straightforward, leads to more complex expressions. 
The factor $\frac{1}{g^{\rho\rho}\left( x^\sigma = \phi^{\left( \sigma \right)} \right)}$ in $S_\lambda$ might look to break the diffeomorphism 
invariance. 
We should note, however, that this factor is not a component of the metric but a function of the scalar fields and therefore this factor 
is consistent with the diffeomorphism invariance.

In $S_\lambda$, $ \lambda^{(\rho)}$'s are Lagrange multiplier fields which give constraints as in (\ref{lambda2}),
\begin{align}
\label{cnstrnt1}
0 = \frac{1}{g^{\rho\rho}\left( x^\sigma = \phi^{\left( \sigma \right)} \right)} g^{\mu\nu} \left( x^\sigma \right)
\partial_\mu \phi^{\left(\rho\right)} \partial_\nu \phi^{\left(\rho\right)} - 1 \, ,
\end{align}
which eliminate ghosts.

By the variation of the action (\ref{acg1}) with respect to the metric $g_{\mu\nu}$, we obtain
\begin{align}
\label{Eqs1}
\mathcal{G}_{\mu\nu} =&\, \frac{1}{2} g_{\mu\nu} \left\{ \frac{1}{2} \sum_{\rho,\sigma=0,1,2,3} A_{\left(\rho\sigma\right)} \left(\phi^{\left(\tau\right)}\right)
g^{\xi\eta} \partial_\xi \phi^{\left(\rho\right)} \partial_\eta \phi^{\left(\sigma\right)} - V\left( \phi^{\left(\rho\right)} \right) \right\}
 - \frac{1}{2} \sum_{\rho,\sigma=0,1,2,3} A_{\left(\rho\sigma\right)} \left(\phi^{\left(\tau\right)}\right)
\partial_\mu \phi^{\left(\rho\right)} \partial_\nu \phi^{\left(\sigma\right)} \nonumber \\
&\, + \frac{1}{2} g_{\mu\nu} \sum_{\rho=0,1,2,3} \lambda^{(\rho)} \left(
\frac{1}{g^{\rho\rho}\left( x^\sigma = \phi^{\left( \sigma \right)} \right)} g^{\xi\eta} \left( x^\sigma \right)
\partial_\xi \phi^{\left(\rho\right)} \partial_\eta \phi^{\left(\rho\right)} - 1\right)
 - \sum_{\rho=0,1,2,3} \frac{\lambda^{(\rho)}}{g^{\rho\rho}\left( x^\sigma = \phi^{\left( \sigma \right)} \right)}
\partial_\mu \phi^{\left(\rho\right)} \partial_\nu \phi^{\left(\rho\right)} \nonumber \\
=&\, \frac{1}{2} g_{\mu\nu} \left\{ \frac{1}{2} \sum_{\rho,\sigma=0,1,2,3} A_{\left(\rho\sigma\right)} \left(\phi^{\left(\tau\right)}\right)
g^{\xi\eta} \partial_\xi \phi^{\left(\rho\right)} \partial_\eta \phi^{\left(\sigma\right)} - V\left( \phi^{\left(\rho\right)} \right) \right\}
 - \frac{1}{2} \sum_{\rho,\sigma=0,1,2,3} A_{\left(\rho\sigma\right)} \left(\phi^{\left(\tau\right)}\right)
\partial_\mu \phi^{\left(\rho\right)} \partial_\nu \phi^{\left(\sigma\right)} \nonumber \\
&\, - \sum_{\rho=0,1,2,3} \frac{\lambda^{(\rho)}}{g^{\rho\rho}\left( x^\sigma = \phi^{\left( \sigma \right)} \right)}
\partial_\mu \phi^{\left(\rho\right)} \partial_\nu \phi^{\left(\rho\right)} \, .
\end{align}
Here we used (\ref{cnstrnt1}).
By multiplying Eq.~(\ref{Eqs1}) with $g^{\mu\nu}$, we find
\begin{align}
\label{Eqs2}
g^{\mu\nu} \mathcal{G}_{\mu\nu} =&\,
\frac{1}{2} \sum_{\rho,\sigma=0,1,2,3} A_{\left(\rho\sigma\right)} \left(\phi^{\left(\tau\right)}\right)
g^{\xi\eta} \partial_\xi \phi^{\left(\rho\right)} \partial_\eta \phi^{\left(\sigma\right)} - 2 V\left( \phi^{\left(\rho\right)} \right)
 - \sum_{\rho=0,1,2,3} \lambda^{(\rho)} \, .
\end{align}
Here we used (\ref{cnstrnt1}), again.
By substituting (\ref{Eqs2}) into (\ref{Eqs1}), we find
\begin{align}
\label{Eqs3}
\sum_{\rho,\sigma=0,1,2,3} A_{\left(\rho\sigma\right)} \left(\phi^{\left(\tau\right)}\right) \partial_\mu \phi^{\left(\rho\right)} \partial_\nu \phi^{\left(\sigma\right)}
=&\, - 2 \mathcal{G}_{\mu\nu}
+ g_{\mu\nu} \left\{ V\left( \phi^{\left(\rho\right)} \right) + \sum_{\rho=0,1,2,3} \lambda^{(\rho)} + g^{\rho\sigma} \mathcal{G}_{\rho\sigma} \right\} \nonumber \\
&\, - 2 \sum_{\rho=0,1,2,3} \frac{\lambda^{(\rho)}}{g^{\rho\rho}\left( x^\sigma = \phi^{\left( \sigma \right)} \right)}
\partial_\mu \phi^{\left(\rho\right)} \partial_\nu \phi^{\left(\rho\right)} \, .
\end{align}
We now identify $\phi^{\left(\rho\right)}=x^\rho$, which is consistent with the constraints in (\ref{cnstrnt1}).
Then (\ref{Eqs3}) can be rewritten as follows,
\begin{align}
\label{Eqs4}
A_{\left(\mu\nu\right)} \left(\phi^{\left(\tau\right)}\right) = - 2 \mathcal{G}_{\mu\nu}
+ g_{\mu\nu} \left\{ V\left( \phi^{\left(\rho\right)} \right) + \sum_{\rho=0,1,2,3} \lambda^{(\rho)} + g^{\rho\sigma} \mathcal{G}_{\rho\sigma} \right\}
 - \frac{2 \lambda^{(\mu)}}{g^{\mu\nu}\left( x^\sigma = \phi^{\left( \sigma \right)} \right)} \delta_{\mu\nu} \, .
\end{align}
Especially, we consider the solution where $\lambda^{(\rho)} =0$.
Then arbitrary geometry given by $g_{\mu\nu}$ and arbitrary function $ V\left( \phi^{\left(\rho\right)} = x^\rho \right)$
can be realized by choosing $A_{\left(\mu\nu\right)} \left(\phi^{\left(\tau\right)}\right)$ as
\begin{align}
\label{Eqs5}
A_{\left(\mu\nu\right)} \left(\phi^{\left(\tau\right)}\right) = - 2 \mathcal{G}_{\mu\nu} \left( x^\sigma = \phi^{\left( \sigma \right)} \right)
+ g_{\mu\nu} \left( x^\sigma = \phi^{\left( \sigma \right)} \right) \left\{ V\left( \phi^{\left(\rho\right)} \right)
+ g^{\rho\sigma} \left( x^\sigma = \phi^{\left( \sigma \right)} \right) \mathcal{G}_{\rho\sigma} \left( x^\sigma = \phi^{\left( \sigma \right)} \right) \right\} \, .
\end{align}
Because the potential $V\left( \phi^{\left(\rho\right)} \right)$ is arbitrary, we may choose $V\left( \phi^{\left(\rho\right)} \right)=0$.
Then the model in this paper can be regarded as a non-linear sigma model whose metric of the target space is given by $A_{\left(\mu\nu\right)} \left(\phi^{\left(\tau\right)}\right)$.
If $A_{\left(\mu\nu\right)}=0$ for a fixed $\mu$ and $\nu=0,1,2,3$ and other all non-vanishing $A_{\left(\rho\sigma\right)}$'s do not depend on
$\phi^{\left(\mu\right)}$ for the fixed $\mu$, we may drop the scalar field $\phi^{\left(\mu\right)}$.

When any eigenvalue of $A_{\left(\mu\nu\right)} \left(\phi^{\left(\tau\right)}\right)$ becomes negative, if there is not the term $S_\lambda$ in (\ref{acg1}),
a ghost appears.
We now check if the constraints (\ref{cnstrnt1}) given by the term $S_\lambda$ in (\ref{acg1}) can eliminate the ghosts.
For this purpose, we may consider the perturbation as in (\ref{pert1}),
\begin{align}
\label{pert1B}
\phi^{\left(\xi\right)}=x^\xi + \delta\phi^{\left(\xi\right)} \, .
\end{align}
For the perturbation $\delta\phi^\xi$, 
the constraints in Eq.~(\ref{cnstrnt1}) give
\begin{align}
\label{pert2B}
0 = 2 g^{\xi\nu}\left(x^\sigma=\phi^{(\sigma)} \right) \partial_\nu \delta \phi^{(\xi)} 
 - \sum_\zeta \delta \phi^{(\zeta)} \partial_\zeta g^{\xi\xi}\left(x^\sigma=\phi^{(\sigma)} \right) \, .
\end{align}
Here, we have not summed the equations with respect to $\xi$. 
In Eq.~(\ref{pert2B}), as long as we consider the perturbation from the background, we may replace 
$\phi^{(\sigma)}$ inside $g^{\xi\nu}\left(x^\sigma=\phi^{(\sigma)} \right)$ and $g^{\xi\xi}\left(x^\sigma=\phi^{(\sigma)} \right)$ with $x^\sigma$. 
For a space-like coordinate $x^\xi$, if we impose $\delta\phi^{(\xi)}=0$ when $\left| x^\xi \right|\to \infty$, 
and for a time-like coordinate $x^\xi$, if we impose $\delta\phi^{(\xi)}=0$ as an initial condition, 
we always find $\delta\phi^{(\xi)}=0$.
Therefore, $\delta\phi^{(\xi)}$ does not propagate, and thus the ghosts do not appear.

In the above discussion, when we consider the perturbation in the fixed background, the scalar fields $\phi^{\left(\xi\right)}$ are not dynamical. 
In the perturbation in (\ref{pert1B}), we do not include the perturbation of the metric. 
In general, some components in the metric may mix with the scalar fields $\phi^{\left(\xi\right)}$ because the scalar fields couple with gravity, of course. 
In fact, Eq.~(\ref{pert2B}) is modified as 
$0 = 2 g^{\xi\nu}\left(x^\sigma \right) \partial_\nu \delta \phi^{(\xi)} 
 - \sum_\zeta \delta \phi^{(\zeta)} \partial_\zeta g^{\xi\xi}\left(x^\sigma \right) - \delta g^{\xi\xi} \left( x^\sigma \right)$. 
The perturbation $\delta g^{\xi\xi}$ includes the scalar mode and the tensor mode with $+$-polarization. 
In Einstein's gravity or even $f(Q)$ gravity, the scalar modes are not dynamical nor do not propagate. 
As a dynamical system, because the scalar mode in the metric is not dynamical, this mode does not have 
conjugate momenta. 
Although there appear conjugate momenta for the scalar fields $\phi^{(\xi)}$'s, they do not propagate by the constraints. 
Therefore even if there is a mixture of the scalar modes, the dynamical modes could not appear. 
Furthermore, although the scalar fields $\phi^{(\xi)}$'s couple with $+$-tensor mode, the scalar perturbation $\delta \phi^{(\xi)}$'s could not mix with the tensor mode.  
At least, it could not generate a new propagating mode although we need more detailed analyses, which could be checked in future 
work including more general gravity theory. 

As in this paper, some extensions of the mimetic gravity have been proposed and investigated
\cite{Takahashi:2017pje, Langlois:2018jdg, Ganz:2018mqi} 
and some instabilities have been also found. 
This may suggest that we need more detailed analyses in order to check the consistencies of the model, which could be also considered 
in the future works.

Finally, we should note that it is straightforward to extend the above model into the model in general $D$ dimensions by using $D$ scalar fields.

\section{$f(Q)$ gravity}
\label{fQ}

Although general relativity is based on the curvature of spacetime, it is possible to construct a theory of gravity based on a geometric quantity different from curvature,
called the torsion or non-metricity tensor. $f(Q)$ gravity theory is a theory in which an action is created from a scalar quantity $Q$.
The quantity $Q$ is constructed from the non-metricity tensor.
If the function $f(Q)$ is a linear function of $Q$, this theory is equivalent to general relativity.
However, the structure of the theory when nonlinear functions are considered as $f(Q)$ is still not well understood.
In this section, we review $f(Q)$ gravity very briefly \cite{Banerjee:2021mqk, Lin:2021uqa}), and consider the possibility of the extensions of this theory
by introducing additional scalar fields as in the previous sections.

\subsection{Brief review}

The general affine connection on a manifold that is both parallelizable and differentiable can be expressed as follows:
\begin{align}
\label{affine}
{\Gamma^\sigma}_{\mu \nu}= {{\tilde \Gamma}^\sigma}_{\mu \nu} + K^\sigma_{\;\mu \nu} + L^\sigma_{\;\mu \nu}\,.
\end{align}
Here $\tilde \Gamma^\sigma_{\;\mu \nu}$ is the Levi-Civita connection given by the metric,
\begin{align}
\label{Levi-Civita}
{{\tilde\Gamma}^\sigma}_{\mu \nu} = \frac{1}{2} g^{\sigma \rho} \left( \partial_\mu g_{\rho \nu} + \partial_\nu g_{\rho \mu}- \partial_\rho g_{\mu \nu}\right)\,.
\end{align}
Furthermore, ${K^\sigma}_{\mu \nu}$ represents the contortion which is defined by using the torsion tensor
${T^\sigma}_{\mu \nu}={\Gamma^\sigma}_{\mu \nu} - {\Gamma^\sigma}_{\nu \mu}$
\begin{align}\label{contortion}
{K^\sigma}_{\mu \nu}= \frac{1}{2} \left( {T^\sigma}_{\mu \nu} + T^{\ \sigma}_{\mu\ \nu} + T^{\ \sigma}_{\nu\ \mu} \right) \, .
\end{align}
Finally, $L^\sigma_{;\mu \nu}$ denotes the deformation and is expressed as follows:
\begin{align}
\label{deformation}
{L^\sigma}_{\mu \nu}= \frac{1}{2} \left( Q^\sigma_{\;\mu \nu} - Q^{\ \sigma}_{\mu\ \nu} - Q^{\ \sigma}_{\nu\ \mu} \right)\,.
\end{align}
Here ${Q^\sigma}_{\mu \nu}$ represents the non-metricity tensor expressed as,
\begin{align}
\label{nonmetricity00}
Q_{\sigma \mu \nu}= \nabla_\sigma g_{\mu \nu}= \partial_\sigma g_{\mu \nu} - {\Gamma^\rho}_{\sigma \mu } g_{\nu \rho} - {\Gamma^\rho}_{\sigma \nu } g_{\mu \rho } \,.
\end{align}
Consequently, the scalar of the non-metricity is defined as:
\begin{align}
\label{non-m scalar}
Q\equiv g^{\mu \nu} \left( {L^\alpha}_{\beta \nu}{L^\beta}_{\mu \alpha} - {L^\beta}_{\alpha \beta} {L^\alpha}_{\mu \nu} \right)
 -Q_{\sigma \mu \nu} P^{\sigma \mu \nu} \,.
\end{align}
Here $P^{\sigma \mu \nu}$ represents the non-metricity conjugate defined by,
\begin{align}
\label{non-m conjugate}
{P^\sigma}_{\mu \nu} \equiv \frac{1}{4} \left\{ - {Q^\sigma}_{\mu \nu} + Q^{\ \sigma}_{\mu\ \nu} + Q^{\ \sigma}_{\nu\ \mu}
+ Q^\sigma g_{\mu \nu}- \tilde{Q}^\sigma g_{\mu \nu} - \frac{1}{2} \left( {\delta^\sigma}_\mu Q_\nu + {\delta^\sigma}_\nu Q_\mu \right) \right\}\, ,
\end{align}
and $Q_\sigma$ and $\tilde{Q}_\sigma$ are defined as
$Q_\sigma \equiv Q^{\ \mu}_{\sigma\ \mu}$ and $\tilde{Q}_\sigma=Q^\mu_{\ \sigma \mu}$.

When both the torsion and the non-metricity equal zero, ${T^\sigma}_{\mu \nu}={Q^\sigma}_{\mu \nu}=0$, the connection ${\Gamma^\sigma}_{\mu \nu}$
reduces to the Levi-Civita connection $ {{\tilde \Gamma}^\sigma}_{\mu \nu}$.
In the Symmetric Teleparallel Equivalent of General Relativity (STEGR), the torsion ${T^\sigma}_{\mu \nu}$ and the curvature $R^\lambda_{\ \mu\rho\nu}$
given by the general affine connection ${\Gamma^\sigma}_{\mu \nu}$ in (\ref{affine}) vanish,
\begin{align}
\label{curvatures}
R^\lambda_{\ \mu\rho\nu} \equiv {\Gamma^\lambda}_{\mu\nu,\rho} - {\Gamma^\lambda}_{\mu\rho,\nu} 
+ {\Gamma^\eta}_{\mu\nu}{\Gamma^\lambda}_{\rho\eta} - {\Gamma^\eta}_{\mu\rho}{\Gamma^\lambda}_{\nu\eta} =0 \, .
\end{align}
A solution of Eq.~(\ref{curvatures}) is given by using four fields $\xi^a$ $\left( a = 0,1,2,3 \right)$ as follows, 
\begin{align}
\label{G1B}
{\Gamma^\rho}_{\mu\nu}=\frac{\partial x^\rho}{\partial \xi^a} \partial_\mu \partial_\nu \xi^a \, .
\end{align}
As we mentioned, there is a problem with the number of degrees of freedom in $f(Q)$ gravity~\cite{Hu:2022anq, DAmbrosio:2023asf, Hu:2023gui}.
Although the functional degrees of freedom in the connection are restricted by the curvature-free and torsion-free conditions, 
it is not so clear which could be valid in the equations given by the variation with respect to the connection. 
In order to avoid this problem, we regard the metric $g_{\mu\nu}$ and $\xi^a$ as independent fields as in \cite{Nojiri:2024zab}. 

Although the STEGR gravity depends on both metric and connection satisfying the above constraints ${T^\sigma}_{\mu \nu}=R^\lambda_{\ \mu\rho\nu}=0$,
the coincident gauge where ${\Gamma^\sigma}_{\mu \nu}=0$ are often used although the bases of the gauge are not so clear but the gauge condition
could correspond to the choice of the coordinate system. 
To realize the coincident gauge, we may use the gauge fixing action $S_\mathrm{gf}$ by introducing the auxiliary fields $\zeta_a$ as follows, 
\begin{align}
\label{Sgf}
S_\mathrm{gf} = \int d^4x \sqrt{-g} \zeta_a \left( \xi^a - x^a \right)\, ,
\end{align}
which violates the general covariance because the coincident gauge corresponds to the coordinate choice. 
We should note that $\zeta_a$ is nothing but the Nakanishi-Lautrup field~\cite{Nakanishi:1966zz, Lautrup:1967zz}. 

In the coincident gauge, the non-metricity tensor in Eq.~(\ref{non-m scalar}) reduces to the following form,
\begin{align}
\label{nonmetricity}
Q_{\sigma \mu \nu}= \nabla_\sigma g_{\mu \nu}= \partial_\sigma g_{\mu \nu} \,.
\end{align}
In the coincident gauge (\ref{nonmetricity}), we have
\begin{align}
\label{Q}
Q= - \frac{1}{4} g^{\alpha\mu} g^{\beta\nu} g^{\gamma\rho} \partial_\alpha g_{\beta\gamma} \partial_\mu g_{\nu\rho}
+ \frac{1}{2} g^{\alpha\mu} g^{\beta\nu} g^{\gamma\rho} \partial_\alpha g_{\beta\gamma} \partial_\rho g_{\nu\mu}
+ \frac{1}{4} g^{\alpha\mu} g^{\beta\gamma} g^{\nu\rho} \partial_\alpha g_{\beta\gamma} \partial_\mu g_{\nu\rho}
 - \frac{1}{2} g^{\alpha\mu} g^{\beta\gamma} g^{\nu\rho} \partial_\alpha g_{\beta\gamma} \partial_\nu g_{\mu\rho} \, .
\end{align}
When we consider the action
\begin{align}
\label{ActionQ}
S_{f(Q)}=\int d^4 x \sqrt{-g} f(Q)\, ,
\end{align}
we find
\begin{align}
\label{eqs}
\mathcal{G}_{\mu\nu} =&\, \frac{1}{\sqrt{-g}} g_{\mu\rho} g_{\nu\sigma}\frac{\delta S_{f(Q)}}{\delta g_{\rho\sigma}} \nonumber \\
=&\, \frac{1}{2} g_{\mu\nu} f(Q)
 - f'(Q) g^{\alpha\beta} g^{\gamma\rho} \left\{ -\frac{1}{4} \partial_\mu g_{\alpha\gamma} \partial_\nu g_{\beta\rho}
 - \frac{1}{2} \partial_\alpha g_{\mu\gamma} \partial_\beta g_{\nu\rho} \right. \nonumber \\
&\, + \frac{1}{2} \left( \partial_\mu g_{\alpha\gamma} \partial_\rho g_{\beta\nu}
+ \partial_\nu g_{\alpha\gamma} \partial_\rho g_{\beta\mu} \right)
+ \frac{1}{2} \partial_\alpha g_{\mu\gamma} \partial_\rho g_{\nu\beta}
+ \frac{1}{4} \partial_\mu g_{\alpha\beta} \partial_\nu g_{\gamma\rho}
+ \frac{1}{2} \partial_\alpha g_{\mu\nu} \partial_\beta g_{\gamma\rho} \nonumber \\
&\, \left. - \frac{1}{4} \left( \partial_\mu g_{\alpha\beta} \partial_\gamma g_{\nu\rho} + \partial_\nu g_{\alpha\beta} \partial_\gamma g_{\mu\rho} \right)
 - \frac{1}{2} \partial_\alpha g_{\mu\nu} \partial_\gamma g_{\beta\rho}
 - \frac{1} {4} \left(\partial_\alpha g_{\gamma\rho} \partial_\mu g_{\beta\nu} + \partial_\alpha g_{\gamma\rho} \partial_\nu g_{\beta\mu} \right)
\right\} \nonumber \\
&\, - \frac{g_{\mu\rho} g_{\nu\sigma}}{\sqrt{-g}} \partial_\alpha \left[ \sqrt{-g} f'(Q) \left\{ - \frac{1}{2} g^{\alpha\beta} g^{\gamma\rho} g^{\sigma\tau} \partial_\beta g_{\gamma\tau}
+ \frac{1}{2} g^{\alpha\beta} g^{\gamma\rho} g^{\sigma\tau} \left( \partial_\tau g_{\gamma\beta} + \partial_\gamma g_{\tau\beta} \right) \right\} \right. \nonumber \\
&\, \left. \left. + \frac{1}{2} g^{\alpha\beta} g^{\rho\sigma} g^{\gamma\tau} \partial_\beta g_{\gamma\tau}
 - \frac{1}{2} g^{\alpha\beta} g^{\rho\sigma} g^{\gamma\tau} \partial_\gamma g_{\beta\tau}
 - \frac{1}{4} \left( g^{\alpha\sigma} g^{\gamma\tau} g^{\beta\rho} + g^{\alpha\rho} g^{\gamma\tau} g^{\beta\sigma}
\right) \partial_\beta g_{\gamma\tau}
\right\} \right]\, .
\end{align}
By combining (\ref{eqs}) with (\ref{Eqs5}), we can construct a model realizing arbitrarily given spacetime. 

We should note that although we chose the coincident gauge, there are equations given by the variation of the total action with respect to the connection. 
In our formulation, the connection is expressed by $\xi^a$. 
Therefore if the matter does not couple with $\xi^a$, we find 
\begin{align}
\label{Xa}
0= \frac{1}{\sqrt{-g}} \frac{\delta \left( S_{f(Q)} + S_\mathrm{gf} \right)}{\delta \xi^a} = X_a + \zeta_a \, , \quad 
X_a \equiv \frac{1}{\sqrt{-g}} \frac{\delta S_{f(Q)}}{\delta \xi^a} \, , 
\end{align}
which can be always solved as $\zeta_a = - X_a$. 
We should note $X_a$ does not vanish in the general spacetime as in the spherically symmetric spacetime or 
the Friedmann-Lema\^{i}tre-Robertson-Walker (FLRW) spacetime. 
Therefore if there is not the gauge fixing action $S_\mathrm{gf}$, Eq.~(\ref{Xa}) is not satisfied $X_a\neq 0$. 
We should note that $X_a$ corresponds to $\nabla_\mu \nabla_\nu \left( \sqrt{-g} f'(Q) {P^{\mu\nu}}_\lambda\right)$. 
In the previous research, the equation $\nabla_\mu \nabla_\nu \left( \sqrt{-g} f' (Q){P^{\mu\nu}}_\alpha\right)=0$ was often imposed and it 
generates some conflictions between the coincident gauge and the spacetime. 
These conflictions occur because the gauge fixing action $S_\mathrm{gf}$ is not included in the total action. 
We should also note that $\frac{1}{\sqrt{-g}} g_{\mu\rho} g_{\nu\sigma}\frac{\delta S_\mathrm{gf}}{\delta g_{\rho\sigma}} = 0$ because $\xi^a=x^a$, 
and therefore the gauge fixing action $S_\mathrm{gf}$ does not contribute to the equations given by the variation with respect to the metric. 

\subsection{Spherically symmetric spacetime}

We now especially consider the construction of the model which realizes arbitrary spherically symmetric but static spacetime.

If we consider a general spherically symmetric and static spacetime by making the metric in (\ref{GBiv}) time-independent, $\nu=\nu(r)$ and $\lambda=\lambda(r)$,
we find
\begin{align}
\label{Qstatic}
Q = - \frac{2 \e^{-2\lambda} \left( 2\nu' r + 1 \right)}{r^2} \, .
\end{align}
We now investigate $\mathcal{G}_{r\vartheta}$,
\begin{align}
\label{rtheta}
\mathcal{G}_{r\vartheta} = \frac{1}{2\tan\vartheta} \frac{d\left( f'(Q)\right)}{dr} \, .
\end{align}
Therefore if we impose a condition $\mathcal{G}_{r\theta}=0$, we find $f''(Q)$ vanishes or $Q$ is a constant.
If $f''(Q)=0$, $f(Q)$ is a linear function of $Q$, which is known to be equivalent to Einstein's gravity.
The condition that $Q$ is a constant gives a strong constraint on the geometry, for example, in the case of Schwarzschild spacetime,
$Q$ is not a constant but we find $Q=-\frac{2}{r^2}$.
Therefore in general $f(Q)$ gravity, the Schwarzschild spacetime cannot be a solution \cite{Wang:2021zaz}.
Even if we use the two-scalar model in Section~\ref{twoscalar}, as clear from Eq.~(\ref{TSBH2B}),
we cannot construct general spherically symmetric spacetime in the framework of $f(Q)$ gravity.
The four-scalar model in Section~\ref{fourscalar} makes the construction of general spherically symmetric spacetime possible
in the $f(Q)$ gravity.

Just for the demonstration, we consider the following specific form of $f(Q)$,
\begin{align}
\label{fq}
f(Q)=\alpha Q+\frac{\beta}{2}Q^2\, .
\end{align}
We just show that we can construct the four-scalar model coupled with $f(Q)$ gravity given by (\ref{fq}), which realizes a spherically symmetric spacetime.

When we describe the spherically symmetric and static spacetime,
instead of using polar coordinates $\left(r,\vartheta,\varphi\right)$, it is often convenient if we use rectangular coordinates $\left(x,y,z\right)=\left(x^i\right)$ $\left(i=1,2,3\right)$.
Because $r=\sqrt{\sum_{i=1,2,3} \left(x^i\right)^2}$, $dr^2 + r^2 \left( d\vartheta^2 + \sin^2 \vartheta d\varphi^2 \right) = \sum_{i=1,2,3} \left( dx^i \right)^2$, $rdr = \sum_{i=1,2,3} x^i dx^i$,
we find the metric~(\ref{GBiv}) with $\nu=\nu(r)$ and $\lambda=\lambda(r)$ can be rewritten as follows,
\begin{align}
\label{ssmetric1}
ds^2 =&\, - \e^{2\nu(r)} dt^2 + \e^{2\lambda(r)} dr^2 + r^2 \left( d\vartheta^2 + \sin^2 \vartheta d\varphi^2 \right) \nonumber \\
=&\, - \e^{2\nu\left(r\sqrt{\sum_{i=1,2,3} \left(x^i\right)^2}\right)} dt^2 + \left( \e^{2\lambda\left(r=\sqrt{\sum_{i=1,2,3} \left(x^i\right)^2}\right)} -1 \right)
\frac{\left( \sum_{i=1,2,3} x^i dx^i \right)^2 }{\sum_{i=1,2,3} \left(x^i\right)^2}
+ \sum_{i=1,2,3} \left( dx^i \right)^2 \nonumber \\
=&\, \sum_{\mu,\nu=t, x, y, z} g_{\mu\nu} dx^\mu dx^\nu \, ,
\end{align}
that is,
\begin{align}
\label{ssmetric2_00}
g_{tt} = - \e^{2\nu\left(r\sqrt{\sum_{i=1,2,3} \left(x^i\right)^2}\right)}\, , \quad
g_{ij} = \left( \e^{2\lambda\left(r=\sqrt{\sum_{k=1,2,3} \left(x^k\right)^2}\right)} -1 \right)
\frac{x^i x^j }{\sum_{k=1,2,3} \left(x^k\right)^2} + \delta_{ij}\, ,
\end{align}
and
\begin{align}
\label{ssmetric2}
g^{tt} = - \e^{-2\nu\left(r\sqrt{\sum_{i=1,2,3} \left(x^i\right)^2}\right)}\, , \quad
g^{ij} = \left( \e^{-2\lambda\left(r=\sqrt{\sum_{k=1,2,3} \left(x^k\right)^2}\right)} -1 \right)
\frac{x^i x^j }{\sum_{k=1,2,3} \left(x^k\right)^2} + \delta^{ij}\, .
\end{align}
By using (\ref{ssmetric2}) and substituting (\ref{fq}) in Eq.~(\ref{Qstatic}), we obtain,
\begin{align}
\label{df22}
A_{\left(t t\right)}=&\, -\frac{2 \e^{2(\nu-2\lambda)}}{r^2} \left\{ r^2 \left[ \left( \alpha + 2r\beta_1 \lambda' +2r\beta_1\nu' \right) \e^{2\lambda}
 -2r\beta_1 \left( \lambda' +\nu' \right) \right]\nu'' +r^2 \left[ \left( \alpha +2r\beta_1\lambda' +2r\beta_1\nu' \right) \e^{2\lambda } \right.\right. \nonumber \\
&\, \left. -2 r\beta_1 \left( \lambda' +\nu' \right) \right] \nu'^2 + \left( 2-\lambda' r \right) \left[ \left(\alpha +2r\beta_1\lambda'
+2 r\beta_1 \nu' \right) \e^{2\lambda} -2r\beta_1 \left( \lambda' +\nu' \right) \right]r \nu' \nonumber \\
&\,\left. - r^2\beta_1 \left( \e^{2\lambda}-1 \right)^2 \left( \lambda'+\nu' \right)^2 \right\} \,,\nonumber\\
A_{\left(i j\right)}=&\, \frac{1}{r^2} \left\{ \left[ 2\left(4\delta_{ij} -5n_in_j \right)r^2\beta_1 \nu'^2-4 r^3 \nu'^3\beta_1 n_i n_j
+ \left[ 4n_i n_j \lambda'^2\beta_1 r^3+ \left(8\delta_{ij} -4n_i n_j\right) \beta_1 r^2\lambda'-4n_i n_j \nu'' \beta_1 r^3\right.\right.\right. \nonumber \\
&\, \left. +2 \left( \delta_{ij}-n_i n_j \right) \left( \alpha+4\beta_1 \right) r \right] \nu'+6n_i n_j\beta_1 \lambda'^2r^2- \left[4n_i n_j \nu''\beta_1 r^2
+2 \left( \delta_{ij}-n_i n_j \right) \left(\alpha-4\beta_1 \right) \right] r\lambda' \nonumber \\
&\, \left. +2\alpha\delta_{ij}-2\alpha n_i n_j \right] {\e^{-2\lambda}}
 -2\beta_1 r \left(\nu' +\lambda' \right) \left( \delta_{ij} -n_i n_j \right)\left(3 \nu'r+2-\lambda'r \right) \e^{-4\lambda}
 - 2n_i n_j\beta_1 r^2 \left( \nu' +\lambda' \right)^2 \e^{2\lambda} \nonumber \\
&\, +4 r^3 \nu'^3\beta_1 n_i n_j+2 \left[ \left( \alpha n_i n_j-\beta_1 \delta_{ij}\right) +3\beta_1 n_i n_j \right] r^2 \nu'^2
 -2r^2\beta_1 \left( \delta_{ij}+n_i n_j \right) \lambda'^2 \nonumber \\
&\, - \left\{ 4n_i n_j r^3 \lambda'^2\beta_1+2r^2 \left[ \left( 2\beta_1\delta_{ij}+ \alpha n_i n_j \right) -2\beta_1 n_i n_j \right] \lambda'
 -4\beta_1 r \left( n_i n_j+ r^2\nu'' n_i n_j-\delta_{ij} \right) \right\}\nu' \nonumber \\
&\, \left.+ \left[ 4n_i n_j \nu''\beta_1 r^2- 4\left( \beta_1\delta_{ij}+\alpha n_i n_j\right)+4\beta_1 n_i n_j \right] r\lambda'
+2 \alpha \left( n_i n_j+ r^2 \nu'' n_i n_j-\delta_{ij} \right)\right\}\,,
\end{align}
where $\beta_1=\frac{\beta}{r^2}$ and $n_i=\frac{x_i} r$.

Then we obtain a model that realizes the spacetime given by the metric~(\ref{GBiv}) with $\nu=\nu(r)$ and $\lambda=\lambda(r)$
in the framework of $f(Q)$ gravity coupled with four scalar fields. 
This could tell that the diffeomorphism invariance is broken in the coincident gauge but the four scalar fields compensate for the invariance. 
This does not mean that the diffeomorphism invariance is recovered but in the different coordinate choices, there are corresponding different theories of 
four scalar fields. 

\section{Summary and discussion}
\label{summaryanddiscussion}

In this paper, we considered the reconstruction problem by using the model with four scalar fields coupled with general gravity theories in (\ref{acg1}).
The model can be a generalization of the two-scalar model (\ref{I8B}) proposed in \cite{Nojiri:2020blr}.
The two-scalar model can realize any given spherically symmetric static/time-dependent spacetime in a wide class of gravity theories as shown in (\ref{Eqs5}).
In the four-scalar model, any given spacetime appears as a solution even if the spacetime does not have a spherical symmetry or any other symmetry.
The two-scalar model includes ghosts, which make the model physically unacceptable.
In the recent works~\cite{Nojiri:2023dvf, Nojiri:2023zlp, Elizalde:2023rds, Nojiri:2023ztz}, however, it has been shown
that the ghosts can be eliminated by constraints (\ref{lambda2}) given by the Lagrange multiplier fields (\ref{lambda1}).
Due to the constraints, the scalar fields become non-dynamical and they do not propagate.
Therefore there is not any sound, which is usually generated by the density fluctuation or oscillation of the fluid.
Although the scalar fields play the role of the effective fluid, the behaviour is rather different from the standard fluid, and the effective
fluid is frozen.
This situation is similar to the mimetic theory in \cite{Chamseddine:2013kea}, where effective dark matter appears from the mimetic
constraint but the effective dark matter has properties different from the standard dark matter.
The mimetic dark matter is non-dynamical and even under gravitational force, it does not collapse.
Even in the four-scalar model in this paper, the ghosts are eliminated by the constraints (\ref{cnstrnt1}).
Therefore, the model in this paper could be the most general extension of the mimetic theory from the effective dark matter
to the effective non-dynamical arbitrary fluid including exotic fluid like a phantom.

We investigated the application of the four-scalar model to $f(Q)$ gravity theory in the coincident gauge.
It is well-known \cite{Banerjee:2021mqk, Lin:2021uqa, Wang:2021zaz, Hohmann:2019fvf, DAmbrosio:2021zpm, Bahamonde:2022zgj, Zhao:2021zab} 
that in the case that $f(Q)$ is a non-linear function of $Q$,
spherically symmetric spacetime cannot be realized in the polar coordinate sysytem except in the case that $Q$ is a constant.
The situation does not change if we use the two-scalar model, as we have shown in (\ref{rtheta})
but in the framework of the four-scalar model in this paper, spherically symmetric spacetime can be realized.

We will now explain why we need four scalar fields to reconstruct the model that realizes a general metric. We will consider a more general scalar-tensor theory with the following action
\begin{align}
\label{st1}
S_\phi = \int d^4x \sqrt{-g} \left\{ \frac{1}{2} \sum_{a,b} A_{(ab)} \left( \phi^{(c)} \right) g^{\mu\nu} \partial_\mu \phi^{(a)} \partial_\nu \phi^{(b)} - V \left( \phi^{(c)} \right)
\right\} \, .
\end{align}
Here we include $N$ scalar fields, $\phi^{(a)}$ $\left( a = 1,2,\cdots,N \right)$.
Note that there is an ambiguity in the redefinition of the scalar fields $\phi^{(a)} \to {\tilde\phi}^{(a)}\left( \phi^{(b)} \right)$, which changes $A_{(ab)}$ by
$A_{(ab)} \left( \phi^{(c)} \right) \to {\tilde A}_{(ab)} \left( {\tilde \phi}^{(c)} \right)\equiv
\sum_{d,f} \frac{\partial \phi^{(d)}}{\partial {\tilde \phi}^{(a)}} {\partial \phi^{(f)}}{\partial {\tilde \phi}^{(b)}} A_{(d f)} \left( \phi^{(h)}\left({\tilde \phi}^{(c)}\right) \right)$. 
This is an analogue of a coordinate transformation.

When $N=1$, through a change of variables, we can set $A_{(ab)}$ to be $\pm 1$. After this adjustment, only one function remains, that is, the potential $V(\phi)$. 
This is a key factor in how the single-scalar model can reconstruct a model that realizes the general FLRW universe \cite{Nojiri:2005pu, Capozziello:2005tf}.

In spherically symmetric spacetime, the metric contains three functions: $g_{tt}$, $g_{tr}=g_{rt}$, and $g_{rr}$. 
Frequently, we can eliminate $g_{tr}$ through a coordinate transformation, as shown in (\ref{GBiv}). 
Therefore, at least two scalar fields are necessary to realize the spherically symmetric spacetime in the framework of the scalar-tensor theory \cite{Nojiri:2020blr}. 
In the two-scalar theory, there are three components of $A_{(ab)} \left( \phi^{(c)} \right)$ and a potential $V \left( \phi^{(c)} \right)$.
Due to the ambiguity of the field redefinition with two functional degrees of freedom, two functional degrees of freedom remain
among the four functional degrees of $A_{(ab)} \left( \phi^{(c)} \right)$ and $V \left( \phi^{(c)} \right)$.

In a general four-dimensional spacetime, the metric possesses ten components, with an additional four redundancies stemming from coordinate transformations. This indicates that, to encompass the general spacetime within the framework of scalar-tensor theory outlined in (\ref{st1}), we require a minimum of four scalar fields. These fields give rise to ten functions $A_{(ab)} \left( \phi^{(c)} \right)$ and a potential $V \left( \phi^{(c)} \right)$, accompanied by four redundancies resulting from the redefinition of scalar fields. 
Therefore, the scalar-tensor theory (\ref{st1}) possesses one additional functional degree of freedom. Consequently, selecting $V=0$ yields a non-linear $\sigma$ model, with its target space being a four-dimensional manifold. 

When the potential in the four-scalar model vanishes, the model becomes a non-linear $\sigma$ model as given in (\ref{Eqs5}).
Therefore when the potential $V$ in (\ref{acg1}), our formulation gives a mapping from the geometry of the spacetime
to the geometry of the target space of the non-linear $\sigma$ model.
We may speculate the meaning more.
When the potential $V$ vanishes, the geometry of the spacetime gives uniquely the geometry of the $\sigma$ model target space.
The inverse is not true because even if we specify the non-linear $\sigma$ model, the spacetime geometry can be determined by solving the partial
differential equations in (\ref{Eqs1}).
This means that the geometry of the spacetime needs, in addition to the geometry of the target space,
the information of the initial or boundary conditions of the scalar fields and the metric, which might be regarded as holographic information.
If the mapping is related to any quantum structure of gravity, it could be interesting.
In the path integral formulation, we need to define the path integration measure.
The measure of the gravity sector might be related to the measure of the non-linear $\sigma$ model although this is merely a speculation.


\begin{thebibliography}{99}

\bibitem{Nojiri:2005pu}
S.~Nojiri and S.~D.~Odintsov,
Gen. Rel. Grav. \textbf{38} (2006), 1285-1304
doi:10.1007/s10714-006-0301-6
[arXiv:hep-th/0506212 [hep-th]].

\bibitem{Capozziello:2005tf}
S.~Capozziello, S.~Nojiri and S.~D.~Odintsov,
Phys. Lett. B \textbf{632} (2006), 597-604
doi:10.1016/j.physletb.2005.11.012
[arXiv:hep-th/0507182 [hep-th]].


\bibitem{Nojiri:2006je}
S.~Nojiri, S.~D.~Odintsov and M.~Sami,
Phys. Rev. D \textbf{74} (2006), 046004
doi:10.1103/PhysRevD.74.046004
[arXiv:hep-th/0605039 [hep-th]].


\bibitem{Nojiri:2005jg}
S.~Nojiri and S.~D.~Odintsov,
Phys. Lett. B \textbf{631} (2005), 1-6
doi:10.1016/j.physletb.2005.10.010
[arXiv:hep-th/0508049 [hep-th]].

\bibitem{Nojiri:2005am}
S.~Nojiri, S.~D.~Odintsov and O.~G.~Gorbunova,
J. Phys. A \textbf{39} (2006), 6627-6634
doi:10.1088/0305-4470/39/21/S62
[arXiv:hep-th/0510183 [hep-th]].

\bibitem{Cognola:2006eg}
G.~Cognola, E.~Elizalde, S.~Nojiri, S.~D.~Odintsov and S.~Zerbini,
Phys. Rev. D \textbf{73} (2006), 084007
doi:10.1103/PhysRevD.73.084007
[arXiv:hep-th/0601008 [hep-th]].

\bibitem{Nojiri:2019dwl}
S.~Nojiri, S.~D.~Odintsov, V.~K.~Oikonomou, N.~Chatzarakis and T.~Paul,
Eur. Phys. J. C \textbf{79} (2019) no.7, 565
doi:10.1140/epjc/s10052-019-7080-1
[arXiv:1907.00403 [gr-qc]].


\bibitem{Nojiri:2006gh}
S.~Nojiri and S.~D.~Odintsov,
Phys. Rev. D \textbf{74} (2006), 086005
doi:10.1103/PhysRevD.74.086005
[arXiv:hep-th/0608008 [hep-th]].

\bibitem{Nojiri:2009kx}
S.~Nojiri, S.~D.~Odintsov and D.~Saez-Gomez,
Phys. Lett. B \textbf{681} (2009), 74-80
doi:10.1016/j.physletb.2009.09.045
[arXiv:0908.1269 [hep-th]].


\bibitem{Sotiriou:2008rp}
T.~P.~Sotiriou and V.~Faraoni,
Rev. Mod. Phys. \textbf{82} (2010), 451-497
doi:10.1103/RevModPhys.82.451
[arXiv:0805.1726 [gr-qc]].

\bibitem{Capozziello:2011et}
S.~Capozziello and M.~De Laurentis,
Phys. Rept. \textbf{509} (2011), 167-321
doi:10.1016/j.physrep.2011.09.003
[arXiv:1108.6266 [gr-qc]].

\bibitem{Nojiri:2010wj}
S.~Nojiri and S.~D.~Odintsov,
Phys. Rept. \textbf{505} (2011), 59-144
doi:10.1016/j.physrep.2011.04.001
[arXiv:1011.0544 [gr-qc]].

\bibitem{Nojiri:2017ncd}
S.~Nojiri, S.~D.~Odintsov and V.~K.~Oikonomou,
Phys. Rept. \textbf{692} (2017), 1-104
doi:10.1016/j.physrep.2017.06.001
[arXiv:1705.11098 [gr-qc]].


\bibitem{Nojiri:2020blr}
S.~Nojiri, S.~D.~Odintsov and V.~Faraoni,
Phys. Rev. D \textbf{103} (2021) no.4, 044055
doi:10.1103/PhysRevD.103.044055 [arXiv:2010.11790 [gr-qc]].


\bibitem{Nashed:2021cfs}
G.~G.~L.~Nashed and S.~Nojiri,
Eur. Phys. J. C \textbf{83} (2023) no.1, 68
doi:10.1140/epjc/s10052-022-11165-4
[arXiv:2112.13391 [gr-qc]].


\bibitem{Kugo:1979gm}
T.~Kugo and I.~Ojima,
Prog. Theor. Phys. Suppl. \textbf{66} (1979), 1-130
doi:10.1143/PTPS.66.1


\bibitem{Chamseddine:2013kea}
A.~H.~Chamseddine and V.~Mukhanov,
JHEP \textbf{11} (2013), 135 doi:10.1007/JHEP11(2013)135
[arXiv:1308.5410 [astro-ph.CO]].


\bibitem{Nojiri:2023dvf}
S.~Nojiri and G.~G.~L.~Nashed,
Phys. Rev. D \textbf{108} (2023) no.12, 124049
doi:10.1103/PhysRevD.108.124049
[arXiv:2309.12379 [hep-th]].


\bibitem{Nojiri:2023zlp}
S.~Nojiri and G.~G.~L.~Nashed,
JCAP \textbf{03} (2024), 023
doi:10.1088/1475-7516/2024/03/023
[arXiv:2310.16068 [gr-qc]].


\bibitem{Elizalde:2023rds}
E.~Elizalde, S.~Nojiri, S.~D.~Odintsov and V.~K.~Oikonomou,
[arXiv:2312.02889 [gr-qc]].


\bibitem{Nojiri:2023ztz}
S.~Nojiri, S.~D.~Odintsov and A.~Sedrakian,
[arXiv:2312.15839 [gr-qc]].


\bibitem{Nojiri:2024dde}
S.~Nojiri, S.~D.~Odintsov and V.~Folomeev,
[arXiv:2401.15868 [gr-qc]].


\bibitem{BeltranJimenez:2017tkd}
J.~Beltr\'an Jim\'enez, L.~Heisenberg and T.~Koivisto,
Phys. Rev. D \textbf{98} (2018) no.4, 044048
doi:10.1103/PhysRevD.98.044048
[arXiv:1710.03116 [gr-qc]].

\bibitem{BeltranJimenez:2019tme}
J.~Beltr\'an Jim\'enez, L.~Heisenberg, T.~S.~Koivisto and S.~Pekar,
Phys. Rev. D \textbf{101} (2020) no.10, 103507
doi:10.1103/PhysRevD.101.103507
[arXiv:1906.10027 [gr-qc]].

\bibitem{Esposito:2021ect}
F.~Esposito, S.~Carloni, R.~Cianci and S.~Vignolo,
Phys. Rev. D \textbf{105} (2022) no.8, 084061
doi:10.1103/PhysRevD.105.084061
[arXiv:2107.14522 [gr-qc]].

\bibitem{DAmbrosio:2021pnd}
F.~D'Ambrosio, L.~Heisenberg and S.~Kuhn,
Class. Quant. Grav. \textbf{39} (2022) no.2, 025013
doi:10.1088/1361-6382/ac3f99
[arXiv:2109.04209 [gr-qc]].

\bibitem{Gadbail:2022jco}
G.~N.~Gadbail, S.~Mandal and P.~K.~Sahoo,
Phys. Lett. B \textbf{835} (2022), 137509
doi:10.1016/j.physletb.2022.137509
[arXiv:2210.09237 [gr-qc]].

\bibitem{Albuquerque:2022eac}
I.~S.~Albuquerque and N.~Frusciante,
Phys. Dark Univ. \textbf{35} (2022), 100980
doi:10.1016/j.dark.2022.100980
[arXiv:2202.04637 [astro-ph.CO]].

\bibitem{Capozziello:2022wgl}
S.~Capozziello and R.~D'Agostino,
Phys. Lett. B \textbf{832} (2022), 137229
doi:10.1016/j.physletb.2022.137229
[arXiv:2204.01015 [gr-qc]].

\bibitem{Khyllep:2022spx}
W.~Khyllep, J.~Dutta, E.~N.~Saridakis and K.~Yesmakhanova,
Phys. Rev. D \textbf{107} (2023) no.4, 044022
doi:10.1103/PhysRevD.107.044022
[arXiv:2207.02610 [gr-qc]].

\bibitem{Anagnostopoulos:2022gej}
F.~K.~Anagnostopoulos, V.~Gakis, E.~N.~Saridakis and S.~Basilakos,
Eur. Phys. J. C \textbf{83} (2023) no.1, 58
doi:10.1140/epjc/s10052-023-11190-x
[arXiv:2205.11445 [gr-qc]].

\bibitem{Mustafa:2021bfs}
G.~Mustafa, Z.~Hassan and P.~K.~Sahoo,
Annals Phys. \textbf{437} (2022), 168751
doi:10.1016/j.aop.2021.168751
[arXiv:2112.15112 [gr-qc]].

\bibitem{Iosifidis:2018diy}
D.~Iosifidis, C.~G.~Tsagas and A.~C.~Petkou,
Phys. Rev. D \textbf{98} (2018) no.10, 104037
doi:10.1103/PhysRevD.98.104037
[arXiv:1809.04992 [gr-qc]].

\bibitem{Dimakis:2022rkd}
N.~Dimakis, A.~Paliathanasis, M.~Roumeliotis and T.~Christodoulakis,
Phys. Rev. D \textbf{106} (2022) no.4, 043509
doi:10.1103/PhysRevD.106.043509
[arXiv:2205.04680 [gr-qc]].

\bibitem{Capozziello:2024vix}
S.~Capozziello, M.~Capriolo and S.~Nojiri,
Phys. Lett. B \textbf{850} (2024), 138510
doi:10.1016/j.physletb.2024.138510
[arXiv:2401.06424 [gr-qc]].


\bibitem{Hu:2022anq}
K.~Hu, T.~Katsuragawa and T.~Qiu,
Phys. Rev. D \textbf{106} (2022) no.4, 044025
doi:10.1103/PhysRevD.106.044025
[arXiv:2204.12826 [gr-qc]].

\bibitem{DAmbrosio:2023asf}
F.~D'Ambrosio, L.~Heisenberg and S.~Zentarra,
Fortsch. Phys. \textbf{71} (2023) no.12, 2300185
doi:10.1002/prop.202300185
[arXiv:2308.02250 [gr-qc]].

\bibitem{Hu:2023gui}
K.~Hu, M.~Yamakoshi, T.~Katsuragawa, S.~Nojiri and T.~Qiu,
Phys. Rev. D \textbf{108} (2023) no.12, 124030
doi:10.1103/PhysRevD.108.124030
[arXiv:2310.15507 [gr-qc]].


\bibitem{Banerjee:2021mqk}
A.~Banerjee, A.~Pradhan, T.~Tangphati and F.~Rahaman,
Eur. Phys. J. C \textbf{81} (2021) no.11, 1031
doi:10.1140/epjc/s10052-021-09854-7
[arXiv:2109.15105 [gr-qc]].

\bibitem{Lin:2021uqa}
R.~H.~Lin and X.~H.~Zhai,
Phys. Rev. D \textbf{103} (2021) no.12, 124001
[erratum: Phys. Rev. D \textbf{106} (2022) no.6, 069902]
doi:10.1103/PhysRevD.103.124001
[arXiv:2105.01484 [gr-qc]].

\bibitem{Wang:2021zaz}
W.~Wang, H.~Chen and T.~Katsuragawa,
Phys. Rev. D \textbf{105} (2022) no.2, 024060
doi:10.1103/PhysRevD.105.024060
[arXiv:2110.13565 [gr-qc]].


\bibitem{Hohmann:2019fvf}
M.~Hohmann,
Symmetry \textbf{12} (2020) no.3, 453
doi:10.3390/sym12030453
[arXiv:1912.12906 [math-ph]].

\bibitem{DAmbrosio:2021zpm}
F.~D'Ambrosio, S.~D.~B.~Fell, L.~Heisenberg and S.~Kuhn,
Phys. Rev. D \textbf{105} (2022) no.2, 024042
doi:10.1103/PhysRevD.105.024042
[arXiv:2109.03174 [gr-qc]].

\bibitem{Bahamonde:2022zgj}
S.~Bahamonde and L.~J\"arv,
Eur. Phys. J. C \textbf{82} (2022) no.10, 963
doi:10.1140/epjc/s10052-022-10922-9
[arXiv:2208.01872 [gr-qc]].

\bibitem{Zhao:2021zab}
D.~Zhao,
Eur. Phys. J. C \textbf{82} (2022) no.4, 303
doi:10.1140/epjc/s10052-022-10266-4
[arXiv:2104.02483 [gr-qc]].


\bibitem{DeFalco:2023twb}
V.~De Falco and S.~Capozziello,
Phys. Rev. D \textbf{108} (2023) no.10, 104030
doi:10.1103/PhysRevD.108.104030
[arXiv:2308.05440 [gr-qc]].


\bibitem{DeFalco:2021klh}
V.~De Falco, E.~Battista, S.~Capozziello and M.~De Laurentis,
Phys. Rev. D \textbf{103} (2021) no.4, 044007
doi:10.1103/PhysRevD.103.044007
[arXiv:2101.04960 [gr-qc]].

\bibitem{DeFalco:2021btn}
V.~De Falco, M.~De Laurentis and S.~Capozziello,
Phys. Rev. D \textbf{104} (2021) no.2, 024053
doi:10.1103/PhysRevD.104.024053
[arXiv:2106.12564 [gr-qc]].


\bibitem{Takahashi:2017pje}
K.~Takahashi and T.~Kobayashi,
JCAP \textbf{11} (2017), 038
doi:10.1088/1475-7516/2017/11/038
[arXiv:1708.02951 [gr-qc]].

\bibitem{Langlois:2018jdg}
D.~Langlois, M.~Mancarella, K.~Noui and F.~Vernizzi,
JCAP \textbf{02} (2019), 036
doi:10.1088/1475-7516/2019/02/036
[arXiv:1802.03394 [gr-qc]].

\bibitem{Ganz:2018mqi}
A.~Ganz, P.~Karmakar, S.~Matarrese and D.~Sorokin,
Phys. Rev. D \textbf{99} (2019) no.6, 064009
doi:10.1103/PhysRevD.99.064009
[arXiv:1812.02667 [gr-qc]].


\bibitem{Nojiri:2024zab}
S.~Nojiri and S.~D.~Odintsov,
Phys. Dark Univ. \textbf{45} (2024), 101538
doi:10.1016/j.dark.2024.101538
[arXiv:2404.18427 [gr-qc]].


\bibitem{Nakanishi:1966zz}
N.~Nakanishi,
Prog. Theor. Phys. \textbf{35} (1966), 1111-1116
doi:10.1143/PTP.35.1111

\bibitem{Lautrup:1967zz}
B.~Lautrup,
NORDITA-214.


\end{thebibliography}
\end{document}